FULL ARTICLE

# Pseudo-real-time retinal layer segmentation for high-resolution adaptive optics optical coherence tomography

Worawee Janpongsri[1] | Joey Huang[1] | Ringo Ng[1] | Daniel J. Wahl[1] | Marinko V. Sarunic[1] | Yifan Jian[*,2,3]

[1] Biomedical Optics Research Group, School of Engineering Science, Simon Fraser University, Burnaby, BC, Canada

[2] Casey Eye Institute, Oregon Health & Science University, Portland, OR, USA

[3] Department of Biomedical Engineering, Oregon Health & Science University, Portland, OR, USA

*Correspondence
Yifan Jian,
Center for Ophthalmic Optics & Lasers,
Casey Eye Institute,
Oregon Health & Science University,
515 S.W. Campus Dr, Portland OR. 97229

Email: jian@ohsu.edu

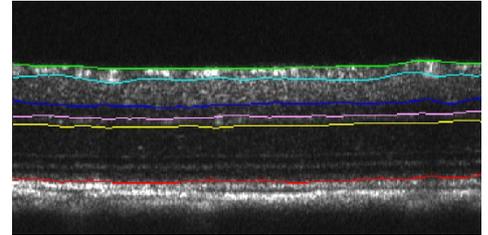

We present a pseudo-real-time retinal layer segmentation for high-resolution Sensorless Adaptive Optics-Optical Coherence Tomography (SAO-OCT). Our pseudo-real-time segmentation method is based on Dijkstra's algorithm that uses the intensity of pixels and the vertical gradient of the image to find the minimum cost in a geometric graph formulation within a limited search region. It segments six retinal layer boundaries in an iterative process according to their order of prominence. The segmentation time is strongly correlated to the number of retinal layers to be segmented. Our program permits *en face* images to be extracted during data acquisition to guide the depth specific focus control and depth dependent aberration correction for high-resolution SAO-OCT systems. The average processing times for our entire pipeline for segmenting six layers in a retinal B-scan of 496x400 pixels and 240x400 pixels are around 25.60 ms and 13.76 ms, respectively. When reducing the number of layers segmented to only two layers, the time required for a 240x400 pixel image is 8.26 ms.

**KEYWORDS**
Retinal layer segmentation, graph search, image processing, pseudo-real-time

## 1 | INTRODUCTION

Optical Coherence Tomography (OCT) is a non-invasive medical imaging technique based on low coherence interferometry, typically using near-infrared light, to capture high-resolution cross-sectional views of biological tissues (e.g., the retina). Ocular imaging with OCT allows ophthalmic clinicians to view and measure the distinctive retinal layer structure to diagnose retinal diseases such as glaucoma and age-related macular degeneration. Also, the diseases related to retinal vessels, such as diabetic retinopathy, can be diagnosed through the retinal vasculature and its hierarchical structure from *en face* OCT Angiography (OCTA) images. OCT technology has continuously improved its acquisition speed; however, due to the complexity of OCT data processing from interferometric fringe data into images, the signal processing is computationally burdensome. Thus, powerful computational resources such as Graphics Processing Units (GPUs) were used to perform the parallelizable aspects of processing interferometric fringes into A-scans, and rendering the resulting volumes [1–8]. Our custom GPU pipeline could perform OCT processing at 2.24 MHz axial scan rate [1] and was also demonstrated for displaying flow contrast *en face* images extracted from the selected depth region on speckle variance OCT (svOCT) angiography [2], [9-10] in real-time.

Rather than perform post-processing analysis, there are ample opportunities and motivations to process the B-scan images to extract additional information as they are acquired. For example, segmentation of the retinal layer boundaries provides an opportunity to perform thickness measurements, which has clinical implications in monitoring glaucoma. In other clinical applications such as computer-assisted surgery, [11–18] high-resolution ocular images can be produced in real time to evaluate dynamic anatomical changes and thus assist clinicians during surgery to potentially increase the rate of success. Moreover, retinal layer segmentation can extract *en* *face* images (taken in the C-scan directions) from the OCT volumes, and real-time retinal layer segmentation can provide effective focus control and direct feedback of aberration correction performance with image-guided AO techniques [19–23]. This last application is the main focus of this report.

Optical aberrations caused by imperfections in the cornea and the intraocular lens reduce the resolution of retinal imaging. Adaptive optics have been integrated with optical retinal imaging to correct ocular aberrations and allow diffraction limited imaging [24]. In particular, our interest is in the combination of AO with OCT [24–31]. The conventional approach to AO uses a Hartmann-Shack Wavefront Sensor (HS-WFS) to detect wavefront distortions and compensate them using a deformable mirror. However, the HS-WFS is sensitive to back-reflections, causing most of the conventional AO to use curved mirrors instead of lenses. We are developing a lens-based sensorless adaptive optics (SAO) approach to correct optical aberrations up to 21$^{st}$ order Zernike polynomials, starting from a defocus [32]. The SAO-OCT can directly evaluate an image quality metric using the extracted *en face* projections to drive the AO correction. Hence, SAO-OCT systems can be used for applications where there are multiple reflecting surfaces from the sample (which could confound a wavefront sensor measurement) or requiring depth resolved aberration correction. While confocal AO scanning laser ophthalmoscope also have depth section capability depending on the confocal pinhole size, AO-SLO is not a volumetric imaging technique, images from only one depth plane can be visualized at a time. With AO-OCT or SAO-OCT, the entire retina can be visualized with a volumetric scan despite limited depth of focus. In addition, combing with real-time retinal layer segmentation, *en face* images generated from any depth plane even if they are out of focus, can be used to drive the SAO correction to shift the focus and sharpen the image.

To extract the thickness measurements and *en face* images from OCT volumes, many segmentation methods have been introduced. Active contours segmentation [33–35] uses an energy formulation; however, it requires a good initialization, and the constraints on the boundaries can cause errors when the retinal layers are in irregular shapes. Besides, the active contours approach also has a high computational cost which is not suitable for real-time applications. Machine learning approaches [36], have been recently introduced, performing retinal segmentation based on learning data representations. They give accurate results; however, they require substantial amount of labelled learning data sets supplied to the networks and it is computationally expensive to train the networks. Automated segmentation methods based on graph theory use pixel intensity and the gradient of the image to find the minimum cost in a geometric graph information for each retinal layer boundary [37–41]. The accuracy and the speed of segmentation based on graph theory depend on the algorithmic implementation used.

To integrate segmentation in real-time OCT imaging applications, a robust segmentation algorithm with low computational cost and low complexity is required. There are several fast retinal layer segmentation attempts for OCT images. Fabritius *et al.* reduced the processing time of segmenting Inner Limiting Membrane (ILM) and Retinal Pigment Epithelium (RPE) by heavily down-sampling each B-scan [42]. The total processing time using a computer with 2.4 GHz CPU to segment a healthy macula volume (1024x320x140) was 6.7 seconds: the average time needed for the ILM was 4.0 seconds and for the RPE was 1.9 seconds. Tian *et al.* developed a faster (~10x) automatic segmentation program called OCTRIMA 3D [40] based on the previous work by Chiu *et al.* [37]. In a follow up paper [43], Tian *et al.* evaluated different segmentation software on a computer with an Intel® Core$^{TM}$ i7-2600 @3.4 GHz CPU. The result showed that OCTRIMA 3D could perform the segmentation of an OCT volume (768 x 496 x 610) in 28 seconds.

In this paper, the main goal is to minimize the processing time of the automated retinal layer program while maintaining the reliability of the segmentation results. We present a pseudo-real-time retinal layer boundaries segmentation program in mice modified from the *Caserel* software [38]. The pseudo-real-time retinal layer segmentation was integrated in our custom GPU pipeline [1] with SAO algorithm [32], which provides focus control and aberration correction for high-resolution real-time visualization of vascular network. The organization of the rest of this report is as follows. The details of the algorithm are presented in Section 2 followed by how we set up the environment for running and testing the algorithm in Section 3. Section 4 shows the results of our segmentation algorithm along with the speed performance and compares the results of AO optimization with and without the segmentation. Discussion about the segmentation performance is covered in Section 5 and the conclusion and future work are presented in Section 6.

## 2 | METHODS

This section describes the implementation of the pseudo-real-time retinal layer segmentation in C/C++ for integration with an OCT acquisition system. The Method description is divided into six subsections: a) image cropping to contain only the area of interest, b) logarithmic scaling and noise reduction, c) layer endpoint initialization and weights calculation, d) ILM and RPE layers segmentation, and e) limiting the search region for the other layers. Figure 1 shows the flowchart outline of the steps for our retinal layer segmentation. Our segmentation software uses graph-cut algorithm to delineate the retinal layers which was first introduced by Chiu *et al* [37]. We perform image cropping on the GPU following the generation of the B-scan image, and then the rest of the program is performed on the CPU in order to gain performance in heterogeneous computing according to the capability of each type of the processor.

In our previous work where we chose to perform a graph-cut on the GPU to segment ILM and RPE of human retina using Push-Relabel Graph-Cut (PR GC) algorithm [44], we found that sometimes the PR GC generates an unwanted region along with the two retinal layers. As a result, we performed connected component labeling (CCL) after segmentation, which is a computational expensive method to identify the two largest connected groups and remove smaller artifacts. According to M. Miao [44], the segmentation pipeline with multiple GPU could only segment ILM and RPE layers on a single OCT B-scan (1024x300x900 pixels) in 57.26 ms which was largely affected by the CCL (12.45 ms). Nevertheless, due to the discontinuity of the pre-built NVIDIA Performance Primitives (NPP) graph-cut APIs after Compute Unified Device Architecture (CUDA) 7.5 version [45], other segmentation methods are needed to replace NPP graph-cut APIs with a better speed performance and without generating artifacts with retinal layers. Although there are several alternatives with both CPU and GPU implementations available for the NPP graph-cut APIs [46–50], these max-flow/min-cut algorithm requires the CCL to remove artifacts which is quite slow.

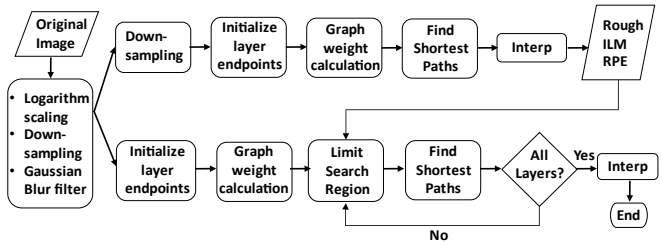

**FIGURE 1** Flowchart of the pseudo-real-time retinal layer segmentation.

Conversely, Dijkstra's algorithm performs a Graph-cut by finding the shortest path or the minimum cost path between two nodes. It produces only a single path without any artifacts. Besides, Dijkstra's algorithm from the Boost Graph Library [51], implemented on the CPU, has a generic interface and can be utilized easily using a header file in C/C++. Therefore, we implemented our retinal layer segmentation program using Dijkstra's algorithm from the Boost Graph Library.

## 2.1 | Pre-processing

In OCT images, particularly when imaging the retina, the data in the axial direction (e.g, the retinal depth) is contained in only a relatively small number of pixels. The pixels that do not contain image information affect the speed performance of the segmentation. In order to decrease the computational cost and make the delineation more reliable, cropping the image to contain only the region of interest (ROI) is necessary. Image cropping was performed on the GPU, immediately after the B-scan image has been generated from the interference signal, and is performed as follows:

Firstly, the average of pixels in each row of the image is computed along the lateral direction:

$$avg(y_j) = \frac{1}{N} \sum_{i=0}^{N-1} f(x_i, y_j) \qquad (1)$$

where $f(x_i,y_j)$ is the grayscale intensity of pixel $(x_i,y_j)$, $i$ and $j$ are the horizontal index and vertical index respectively, and $N$ is the width of the image.

Secondly, we generate a histogram on the row average where the number of bins is chosen to be ten and the size of the bins is calculated as:

$$bin\_size = \frac{\max(row\_avg) - \min(row\_avg) + 1}{number\_of\_bins} \qquad (2)$$

After the histogram is computed, the value of the bin that contains the most common elements is selected as the threshold. Empirically, the first and the last indices of the rows for which the average values are greater than the threshold correspond to the position where the retinal structure begins (ILM layer) and ends (RPE layer), respectively. In order to include all the retinal characteristics in the ROI, the indices where the ROI begins, and ends are calculated as:

$$ROI_{begin\_index} = i - offset \times height_{image} \qquad (3)$$

and

$$ROI_{begin\_index} = i - offset \times height_{image} \qquad (4)$$

where $i$ and $j$ are the first and the last indices where the value of the row average are greater than the threshold. In this study, we set the *offset* = 0.1. We only applied the cropping step on images with height greater than 200 pixels; otherwise, we skip this step because limiting the ROI may crop out some important retinal characteristics for segmentation. Figure 2 shows the original B-scan, its cropped image and the Gaussian blurred image. In addition, we implemented this step on the GPU for faster parallel operations performance on a large set of data.

Before segmenting the OCT retinal images, we employed a logarithm operation on each pixel on the image to compress the dynamic range of the image and enhance low intensity pixel values. Then we down-sampled the cropped image by a factor of two and referred to this as a *resized image*. Removing the noise before further processing the images is essential as the noise could affect the quality of the automated segmentation. The most common noise in OCT images is the speckle noise which is produced by constructive/destructive interference. Speckle noise appears as white and black intensity fluctuations and can be reduced in appearance by applying a Gaussian blur filter. After down-sampling the images, we applied a Gaussian blur filter on the resized image (Figure 2 c).

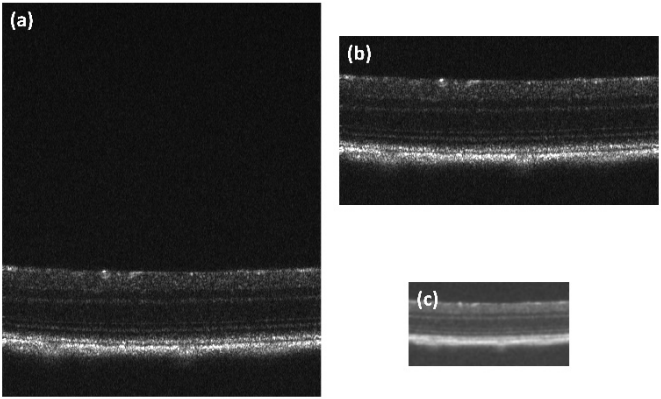

**FIGURE 2** (a) Original B-scans (496x400 pixels), (b) its cropped image (210x400 pixels) and (c) Gaussian blurred image (105x200).

## 2.2 | Layer endpoint initialization and weights calculation

Each image is considered as a graph of nodes, in which a node equates to a pixel on the image having edges connecting to other nodes [49]. A graph may consist of multiple layered structures, and segmenting a particular layer requires the selection of the start and the end nodes. The start and the end nodes are automatically initialized by assuming that the retinal layers to be segmented extend across the entire width of the image. One vertical column is added to each side of the Gaussian blurred image, and they are assigned with zero values. The start node is the left top corner pixel and the end node is the right bottom pixel. These additional columns are removed after the segmentation is completed.

In the retinal images, the foreground is defined as the retinal layers and the background as the vitreous and posterior chamber. The transition in pixel intensity from the background to the foreground is large so a graph can be simply constructed based on calculating the vertical gradient of the image. Figure 3 shows the two gradient images (the positive and negative gradient) of size 105x202 pixels. Both of them are generated because some retinal layer boundaries, such as the vitreous/ILM appears to have a darker layer above a brighter layer, whereas other boundaries, such as the Nerve Fiber Layer/Ganglion Cell Layer (NFL/GCL), have a lighter layer above a darker layer.

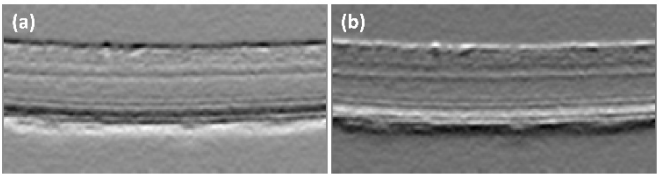

**FIGURE 3** (a) The gradient and (b) the negative gradient images.

The weight of the edges usually represents the geometric distance and/or the intensity difference between the neighboring pixels. However, in OCT retinal images, the features of interest have a smooth transition between neighboring pixels and each pixel is only connected with its eight nearest neighboring pixels and disconnected with other nodes. Hence, the weight of the edges is a function of the intensity difference between the neighboring pixels. Because the retinal layers in OCT images are horizontal structures distinguishable by a vertical change in pixel intensity, the weights are calculated based on the vertical gradient. The formula used in this method for calculating the weights is:

$$w_{ab} = 2 - (g_a - g_b) + w_{min} \qquad (5)$$

where $w_{ab}$ is the weight assigned to the edge connecting nodes $a$ and $b$, $g_a$ is the vertical gradient of the image at node $a$, $g_b$ is the vertical gradient of the image at node $b$ and $w_{min}$ is the minimum weight in the graph (1E-5).

The weights are also calculated based on the directionality of the gradient. As the result, we have two sparse adjacency matrices of intensity difference graph weights of size [MN x MN] with MNC filled entries where [M x N] is the image size and C is the number of the nearest neighbors (in this case is eight). The light-to-dark sparse adjacency matrix of graph weights was created using the gradient whereas the dark-to-light sparse adjacency matrix of graph weights used the negative gradient.- As mentioned above, we added one additional column on each side of the image, we assigned the weight values in those additional columns to be $w_{min}$ so the shortest path calculation would not be affected by the additional columns. The edge weight of zero indicates that the two nodes are unconnected.

## 2.3 | ILM and RPE layer segmentation

The retinal layers are segmented in an iterative process according to their order of prominence. The ILM and RPE layers are segmented first due to their high contrast in pixel intensity relative to the background. To begin the retinal segmentation, the Gaussian blurred image is again resized by a factor of two to roughly segment the ILM and RPE layers. This twice down-sampled image is referred to as the *rough image*. Then we produce the gradient image and the dark-to-light sparse adjacency matrix of graph weights for the rough image. A ROI matrix of the same size as the rough image with two additional columns is generated and each pixel value of the ROI is set to 1 if the corresponding pixel on the rough image is greater than the mean value of the rough image, otherwise zero. We set the value of dark-to-light sparse adjacency matrix of graph weights to zero where the ROI is zero, otherwise the value is not changed.

The procedure described above helps with indicating the region to find the shortest path because zero edge weight means unconnected nodes. Next, we used Dijkstra's algorithm and the dark-to-light sparse adjacency matrix of graph weights to find the shortest paths. After the first layer was segmented, we set the pixels of the first found layer on the ROI matrix to zero in order to segment the second rough layer. We iterated the process to find the second rough layer, setting the value of the dark-to-light sparse adjacency matrix of graph weights to zero where the ROI is zero; otherwise the same value is kept.

After the two rough layers are found, both of the rough layers were interpolated to have the same dimensions as the resized image. Next, we set the layer where the mean value of the y-coordinate is smaller to be the ILM and the other layer to be the RPE. Then we use the ROI matrix of the same dimensions as the resized image with the additional columns to find the precise ILM and RPE layers. First, we segment the ILM by setting the region of the ROI matrix near the rough ILM layer to one where the rest are zero. Also, we change the value of the dark-to-light sparse adjacency matrix of graph weights to zero where the ROI matrix is zero. Then again, we use Dijkstra's algorithm and the dark-to-light sparse adjacency matrix of graph weights to find the precise ILM layer. The precise RPE is found in the same manner as the ILM. Figure 4 shows the ROI images for finding the rough layers and the precise ILM and RPE layers on the resized image (with two additional columns). The ROI images are 53x102 pixels whereas the resized image is 105x202 pixels.

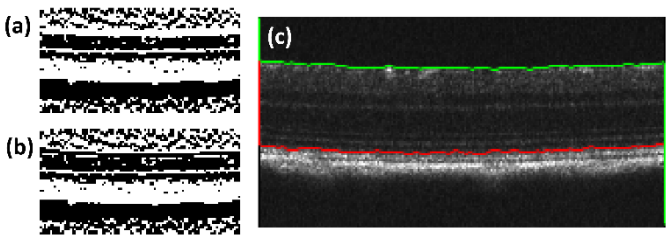

**FIGURE 4** (a) ROI for the first rough layer (b) ROI for the second rough layer and (c) precise ILM and RPE layers on the resized image with two additional columns.

### 2.4 | Limiting the search region for the other layers

As mentioned earlier, due to the hyper-reflectivity of the ILM and RPE layers, they are easily segmented. In contrast, the remaining layers are not as prominent because their characteristics (relative intensity) are similar. In order to correctly segment the targeted layer, the search region should be limited such that the irrelevant features are excluded. This exclusion is accomplished by setting the weight of the non-targeted features to zero before segmenting the graph using Dijkstra's algorithm. The search space of each layer is selected based on the previously segmented layers. The order of layer boundaries to be segmented is Inner Nuclear Layer/Outer Plexiform Layer (INL/OPL), NFL/GCL, Inner Plexiform Layer/Inner Nuclear Layer (IPL/INL), and Outer Plexiform Layer/Outer Nuclear Layer (OPL/ONL). Table 1 shows the upper and lower boundaries and the sparse adjacency matrix of graph weights for segmentation. Each boundary requires two previously segmented boundaries to be the upper and the lower bounds to limit the search region.

**TABLE 1** Upper and lower bounds, and the sparse adjacency (S.A.) matrix used for performing graph search for a particular layer.

| Retinal layer | Upper Bound | Lower Bound | S.A. Matrix |
| --- | --- | --- | --- |
| INL/OPL | ILM | RPE | Dark-to-light |
| NFL/GCL | ILM | INL/OPL | Light-to-dark |
| IPL/INL | NFL/GCL | INL/OPL | Light-to-dark |
| OPL/ONL | INL/OPL | RPE | Light-to-dark |

Once all the retinal layer boundaries are segmented, the additional column on each side of the image is removed, leaving the accurate six retinal layer boundaries. Nonetheless, these retinal layers are not the same size as the original input image, and they need to be interpolated, smoothed and offset adjusted based on the cropped background to correctly delineate these features on the original uncropped image. Figure 5 shows representative result of the retinal layer boundaries delineated on mouse cross-sectional retinal image.

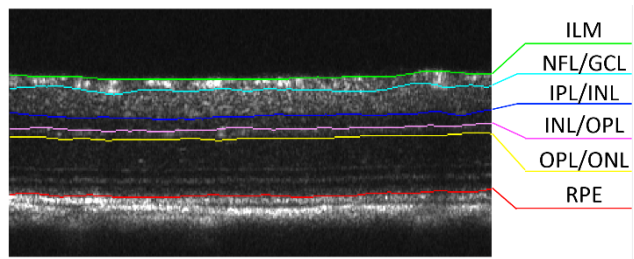

**FIGURE 5** The SD-OCT mouse B-scan with the retinal layer boundaries segmented by the pseudo-real-time segmentation.

## 3 | EXPERIMENTAL SETUP

This section provides an overview of our SAO-OCT imaging system for small animals [32] and our processing and display program for the OCT retinal images [1]. The environment used for implementing and testing our retinal layer segmentation program is described, and the imaging results along with speed performance are presented.

### 3.1 | Overview of our OCT processing and display program

Real-time application of the OCT requires high throughput and low overhead (latency). In this research, we used the parallelization strategies introduced by Jian *et al.* [1] to accelerate OCT processing. To fully utilize the PCIe bandwidth, we transferred the interferometric data from the host to the device as a batch rather than a single frame. In order to hide memory transfer latency, the memory transfer from the host to the device and the data processing on the device was implemented using two CUDA streams concurrently; one to transfer the data processing on the device and another to process the interferometric fringe data on the device. While the small batch of the interferometric fringe data was being transferred from the host to the device by the transfer stream, the previous batch that is already in the device is simultaneously being processed by the kernel stream. These two CUDA streams, which are executed simultaneously, are synchronized after processing each batch. The original implementation by Jian *et al*. [1] demonstrated a high throughput; however, it suffered from a large latency as its processing pipeline was completely asynchronous with the acquisition. In this project, we improved the program latency by synchronizing the data acquisition and processing at the batch level.

### 3.2 | SAO-OCT imaging system for small animals

Our imaging system is a multi-modal small animal retina imaging system that includes OCT, OCT-A, confocal scanning laser ophthalmoscopy (cSLO), and fluorescence detection [32]. It is a compact lens-based system incorporating the SAO technique to correct the optical aberrations instead of using a wavefront sensor to measure the aberrations. For this project, we only used the modified OCT subsystem for mouse imaging. The OCT subsystem in this project used a light source with a central wavelength of 810 nm and a bandwidth of 100 nm. We also integrated our retinal layer segmentation program in our OCT processing and display program along with the SAO–OCT imaging system. The retinal layers were segmented on the cross-sectional images and these results were used to project *en face* images from the selected retinal layers. Then the SAO-OCT used the extracted *en face* projections as the input of its merit function defined as the sum of the intensity squared of each pixel of the *en face* image to drive the optimization algorithm. The OCT acquisition modality provided a 100 kHz A-scan rate for retina imaging and the OCT volumes are acquired with user selected dimensions. Two B-scans were acquired at the same lateral location to generate an OCT-A B-scan image [2]. Our retinal segmentation algorithm was implemented in C/C++ using Microsoft Visual Studio 2013, running on a personal computer with the CPU of Intel® Core™ i9-9900K CPU @3.6 GHz with a Graphics Processing Unit of NVIDIA GeForce RTX 2060. In this study, we used the mouse SD-OCT volume datasets of different dimensions (as indicated below), but each dataset contained 800 frames.

The mouse imaging experiments were performed under protocols compliant to the Canadian Council on Animal Care, and with the approval of the University Animal Care Committee at Simon Fraser University.

## 4 | RESULTS

For the speed performance, we compared our segmentation in C++ with the modified *Caserel* software using MATLAB R2019a. Since our OCTViewer processes the interferometric fringe data into the processed floating-point data, we modified the *Caserel* software to read these types of input. We also modified the resize scale of rough layers to 0.5 instead of 0.2 because if the image size is too small, the path of the rough second layer cannot be found due to the broken path on the ROI. Table 2 shows the accumulated average speed performance of our segmentation up to the specified layer using our C++ implementation and the modified version of *Caserel* software. The order of retinal layers to be segmented is rough ILM and RPE, precise ILM, precise RPE, INL/OPL, NFL/GCL, IPL/INL and OPL/ONL. On average, our implementation can achieve a ~5x acceleration compared to the Caserel software. We also compared the segmentation accuracy of the two implementations on 800 B-scans, the mean pixel difference in retinal layer segmentation between the two methods is negligible (ILM, 0.89 ± 0.19; RPE, 2.05 ± 0.32; INL/OPL, 0.12 ± 0.97; NFL/GCL, 0.24 ± 0.23; IPL/INL, 0.07 ± 0.76; OPL/ONL, 0.11 ± 0.93).

**TABLE 2** Speed performance of C/C++ segmentation and *Caserel* in milliseconds on different image sizes.

| Retinal Layer | 496x400 pixels | | 240x300 pixels | |
|---|---|---|---|---|
| | Caserel | C++ | Caserel | C++ |
| Rough | 51.8 ± 8.2 | 9.1 ± 0.3 | 21.1 ± 7.5 | 5.3 ± 0.4 |
| ILM | 64.9 ± 8.4 | 11.4 ± 0.5 | 28.2 ± 7.7 | 6.8 ± 0.4 |
| RPE | 77.4 ± 8.7 | 13.5 ± 0.5 | 35.2 ± 7.8 | 8.2 ± 0.4 |
| INL/OPL | 91.4 ± 10.1 | 18.8 ± 0.4 | 43.9 ± 8.2 | 10.6 ± 0.6 |
| NFL/GCL | 104.6 ± 11.1 | 20.9 ± 0.5 | 51.4 ± 8.7 | 11.7 ± 0.6 |
| IPL/INL | 117.4 ± 11.7 | 22.4 ± 0.5 | 58.2 ± 8.9 | 12.3 ± 0.8 |
| OPL/ONL | 130.5 ± 13.2 | 25.6 ± 0.6 | 66.2 ± 9.3 | 13.7 ± 0.7 |

Figure 6 compares the results of the intensity and speckle variance *en face* images when using static (fixed) depth locations to generate the *en face* images versus using the pseudo-real-time retinal layer segmentation. With the static user-selected depth option, the operator selects two horizontal lines on the B-scan image. Figure 6 (a) shows a representative B-scan with the static user-selected depth at NFL layer and its corresponding *en face* images. Figure 6 (b) shows a representative B-scan with the pseudo-real-time retinal layer segmentation results of ILM and NFL/GCL layers with its OCT and OCT-A *en face* images. Although the images are similar, the top row (with the segmentation off) is missing a vasculature feature near the top edge of the OCTA image.

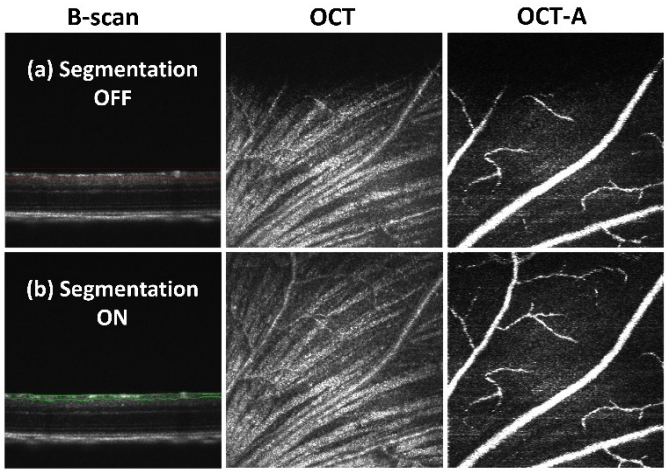

**FIGURE 6** Structural and OCTA NFL *en face* images using static user-selected depth (segmentation OFF) of retinal layers and using the pseudo-real-time retinal layer segmentation (segmentation ON).

Mouse retinal axial motion during imaging is typically on the order of 4.34 pixels during imaging due to breathing and related motion. Hence, without tracking of the retina position, the region of interest may shift in and out of the bounding box. Figure 7 shows a representative case of when the static user-selected depth option retinal layers option failed to detect the retinal layer of interest during data acquisition due to motion in *vivo* mouse.

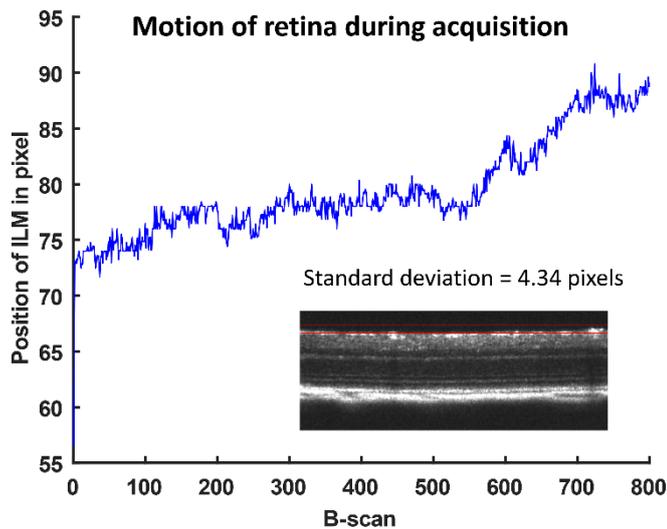

**FIGURE 7** Axial motion of the retina during acquisition causes failure in detecting NFL layer using two fixed horizontal lines and the plot showing the motion of the ILM position (in pixel) during acquisition.

The results for NFL imaging with OCT and OCTA (e.g., not aberration corrected), SAO-OCT, and SAO-OCT-A visualization with and without our pseudo-real-time segmentation are shown in Figure 8. Without our segmentation, an operator selected two straight lines within the region of interest in order to obtain *en face* images. Figure 8 shows *en face* OCT images in (a) and OCT-A images in (b) with and without the segmentation and with and without SAO optimization. Figure 8 (c) shows a representative plot of the SAO merit function value during optimization with and without the segmentation.

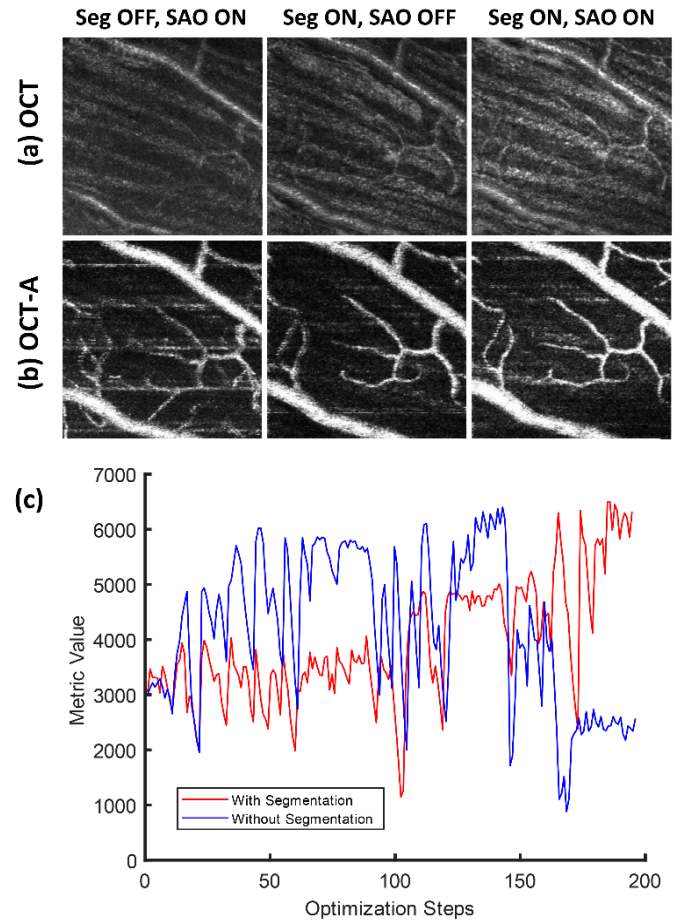

**FIGURE 8** (a) and (b) *En face* images of NFL layer with and without segmentation and SAO. (c) Image quality of each step in the SAO optimization with segmentation (red plot) and without segmentation (blue plot).

Figure 9 shows the segmentation results of different retinal layers from different mice. The left column on each row shows the cross-sectional mouse retina images with two segmented retinal layers. The middle and the right columns displays the intensity and speckle variance *en face* images generated from the segmented layers shown on the corresponding B-scan. Figure 9 (a) shows the results from segmenting the ILM (green line) and RPE layers (red line). In Figure 9 (b), the *en face* images are generated from the OPL layer (the green line is the boundary between INL and OPL and the red line is the boundary between the OPL and ONL). Figure 9 (c) shows the segmentation results from the boundary between NFL/GCL (green line) and IPL/INL (red line) layers, whereas Figure 9 (d) displays the segmented layers of IPL/INL (green line) and OPL/ONL (red line)

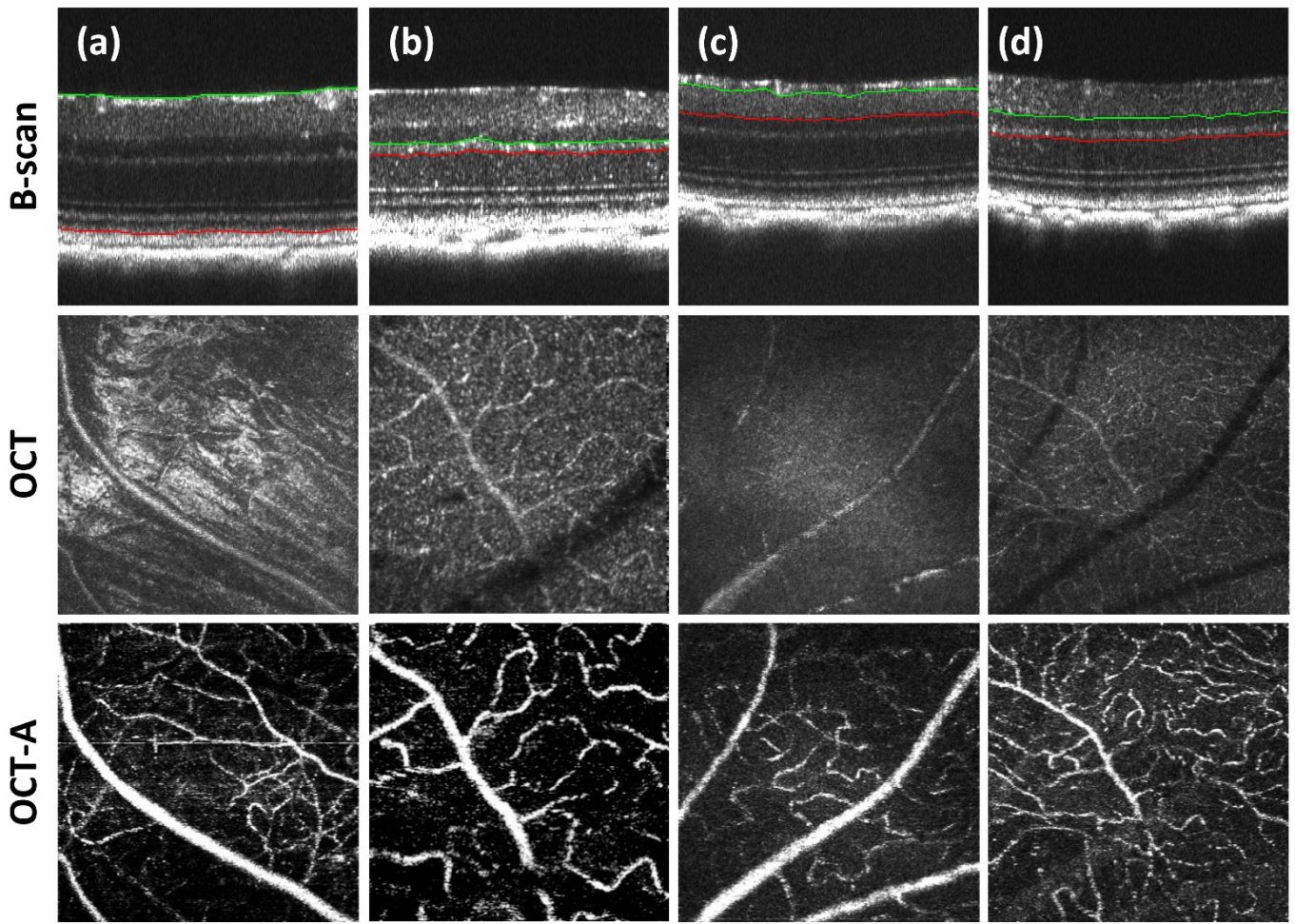

**FIGURE 9** Mouse B-scan images with selected depths of interest with OCT and OCT-A enface images.

## 5 | DISCUSSION

The main objective of this project is to implement a pseudo-real-time retinal layer segmentation program that provides *en face* images generation based on anatomical retinal layer structures. Our retinal layer segmentation method was based graph-cut that use pixel intensity and the gradient of the image, this algorithm has been proven to be robust and reliable for both diseased and non-diseased retina, and it has been used in multiple clinical studies [37–41], [52–55]. Although in this report the *in vivo* experiments performed with wild type mice without retinal pathologies, we expect that our retinal segmentation method can achieve similar performance in segmenting retinas with diseases and abnormalities.

We employed heterogeneous computing to gain the performance based on the nature of each task. CPUs are good at performing complex tasks, and in the context of Graph-cut search, are more reliable in terms of global convergence. In contrast, GPUs are optimized for performing tasks of lesser complexity that can be broken down into smaller independent parts that need less or no communication among tasks. We parallelized the code for cropping the image to contain only the retinal structure and to remove the redundant data by implementing it into CUDA for NVIDIA GPU. Removing redundant data helps Dijkstra's algorithm, which is performed on the CPU due to its capability to perform complex tasks, and to search for the shortest path faster because of the smaller (resized) image size. Moreover, our retinal layer segmentation program utilizes arrays and Intel® Math Kernel Library (MKL) rather than vectors and their operations. Arrays take less time to access their elements because of their contiguous property, and permit access to the elements efficiently with a constant time irrespective of the element location. Since the array is a fixed size data structure, and all elements must be of the same type, hence, it is type safe and the most efficient in terms of speed and performance.

The specifications of the segmentation time requirements (including the OCT signal processing time) were set by the acquisition system parameters. The image acquisition system used in this report provided a 100 kHz A-line scan rate, and completed a B-scan consisting of 400 A-lines in 4 ms and a B-scan consisting of 300 A-lines in 3 ms, corresponding to B-scan frame rates of ~250-333Hz. In order to maximize the data transfer speed across the PCIe bus, the B-scans are acquired and transferred in batches. Therefore, the combination of our CUDA and segmentation pipelines (the processing pipeline) must guarantee the segmentation outcome of the current batch before the next batch is completely acquired. Our segmentation using C/C++ showed a significant improvement in the processing time of a B-scan by more than 74% compared to the processing speed of *Caserel*, and is able to perform the rough segmentation on each (cropped) B-scan at the rate that it is acquired. We found that the lateral movement of the retinal layers within the acquisition time of a batch is negligible; this is particular true for OCT-A where multiple B-scans are acquired at the same position to generate flow contrast. Thus, we reduced the number of frames to be segmented by applying the full resolution segmentation using graph-cuts to the first frame of each batch and we applied segmentation results of the first frame in each batch to all frames in that batch. If we chose to segment only the ILM and RPE layers and chose a batch to contain 4 frames, the acquisition time is about 16 ms for the image with width of 400 pixels and 12 ms for the image with a width of 300 pixels. As shown in Table 2, the execution time to segment the rough and the precise ILM and RPE on an image of size 496x400 takes about 13.57 ms, and on an image of size 240x300 takes about 8.26 ms. As a result, our segmentation pipeline can run at least at 62.5 Hz which is faster than a video rate of 60 Hz. However, we must consider the OCT acquisition and signal processing time along with the image segmentation time to respond within the batch acquisition time. Thus, we chose a batch size to contain 6 frames which extends the acquisition time to 24 ms for the images with 400 pixels width and 18 ms for images with 300 pixels width to include the OCT signal acquisition and processing time. This feature would provide axial tracking and extraction of the shape of the retina for visualization during acquisition. If we wanted to segment a specific retinal layer, we could offset the width between ILM and RPE layers to generate an *en face* image that is close to the anatomical shape of that layer. The segmentation of additional layers would require larger batch sizes. For example, we could also segment all six retinal layers and choose the batch size to be 10 frames to generate the results before the acquisition time of 40 ms and 30 ms for the images where the width is 400 pixels and 300 pixels, respectively. In addition, in most cases, we do not need all six retinal layers delineated on the image while performing a real-time application. Instead, we often need to segment a particular layer by using two segmented retinal layer boundaries. For instance, the boundaries between INL/OPL and OPL/ONL are needed for segmenting the OPL layer. We could speed up our segmentation pipeline by segmenting only the rough ILM and RPE layers and use these rough layers to limit the search region to segment INL/OPL and again use

INL/OPL and rough RPE to get OPL/ONL. Similar to the NFL layer, we could segment the rough layers and use the rough layers to segment INL/OPL and again use rough ILM and INL/OPL to segment NFL/GCL.

Alternative computational hardware could also be used for OCT signal processing and retinal layer segmentation. Our approach focused on the use of GPU and CPU because of the ease of implementation, ability for rapid changes to accommodate emerging applications, and for integration with relatively complex control of adaptive optics systems. Future work may involve the use of FPGAs. The OCT processing pipeline has been developed on FPGA previously [56–60]. With its advantage of the execution time and power consumption, an FPGA platform could also be a suitable option for speeding up the retinal layer segmentation. To the best of our knowledge, there is no work of retinal layer segmentation that uses FPGA; however, there are some works on image segmentation using graph-cut on FPGA [61-62].

In this report, we implemented a pseudo-real-time 2D Graph-cut for retinal layer segmentation which was integrated into a complete OCT acquisition and processing software. We demonstrated the system performance for generating layer segmentations to guide an image-based SAO optimization [63–65]. With the ability to segment the retinal layers, our processing program created anatomically correct *en face* images by using a maximum intensity projection (MIP) between the two selected retinal layers for guiding the SAO optimization in real-time. The use of a static user-selected depth region of interest is not ideal to detect specific retinal layers because, for *in vivo* mouse imaging, the retinal position moves during the acquisition as shown in Figure 8. Figure 6 shows that the top of the NFL *en face* images are dark because during the acquisition, the retina moved out of the depth that is on focus. In contrast, the pseudo-real-time retinal segmentation generated *en face* images contained the correct retinal features because the segmented lines followed the motion of the retina *in vivo*. The pseudo-real-time retinal layer segmentation tracks the layer of interest even though there is motion during acquisition which leads to a better *en face* image as the input for the SAO optimization as shown in Figure 8. Figure 8 (c) shows the merit plots with and without our segmentation for SAO optimization of 18 modes in total. The merit plot with segmentation shows a better improvement in the image quality after each mode is optimized than the merit plot without the segmentation.

## 6 | CONCLUSION

We have utilized consumer grade PC to control the acquisition of real-time OCT signals and perform image processing. A GPU was used for OCT processing and for generating the B-scan images. The CPU was used for the processes such as shortest-path graph search to segment the retinal layers on the acquisition computer and extracting a depth resolved layer from the volumetric data as it gets acquired. We employed down-sampling and parallel processing to improve the speed of the application and as a result our retinal segmentation program can be used as a pseudo-real-time application. Segmentation of the retinal layers permits OCT and OCTA *en face* images to be extracted during data acquisition and used to guide the depth specific focus control for high-solution OCT systems. We demonstrated our retinal segmentation program in real-time in *vivo* mouse retinal imaging, and it can be applied in human retinal imaging in the future.

**ACKNOWLEDGMENTS**

Funding for this work was generously provided by the National Sciences and Engineering Research Council of Canada (NSERC), Canadian Institutes of Health Research (CIHR), the Alzheimer Society of Canada, the Pacific Alzheimer Research Foundation (PARF), the Michael Smith Foundation for Health Research (MSFHR), Unrestricted Grant from Research to Prevent Blindness, and National Institute of Health P30 EY010572 Core Grant.